\documentclass[a4paper,12pt]{article}
\pdfoutput=1
\usepackage{feynmp-auto,expdlist}
\usepackage{amsmath, amsfonts}
\usepackage{graphicx,arydshln}
\usepackage{enumerate}
\usepackage{hyperref}
\usepackage{latexsym}
\usepackage{dsfont}
\usepackage{hepnicenames}
\usepackage{enumerate}
\usepackage{soul}
\usepackage[normalem]{ulem}
\usepackage{comment}

\newcommand{\footnoten}[1]{}

\usepackage[font={small},textfont={it}]{caption} % make captions small

\newcommand{\cW}{c_{\rm W}}
\newcommand{\sW}{s_{\rm W}}

\usepackage{units}

\def\permille{\ensuremath{{}^\text{o}\mkern-5mu/\mkern-3mu_\text{oo}}}
\usepackage{mathrsfs,graphicx,rotating,amsmath,amsfonts,mathtools,booktabs,amssymb,wasysym}
\usepackage{hyperref}\usepackage{slashed}
\usepackage[nosort]{cite}
\usepackage[table,xcdraw,dvipsnames]{xcolor}
\usepackage{bm}
\usepackage{multirow,multicol}
\hypersetup{colorlinks,bookmarksopen,bookmarksnumbered,
	linkcolor=blus,pdfstartview=FitH,urlcolor=rossos,citecolor=verde}
\allowdisplaybreaks

\renewcommand{\[}{\left[}

\def\Lag{\mathscr{L}}

\newcommand{\mio}[1]{}

\newcommand{\med}[1]{\langle #1\rangle}

\def\bpm{\begin{pmatrix}}
	\def\epm{\end{pmatrix}}

\usepackage{mathrsfs}

\newcommand{\fig}[1]{~\ref{fig:#1}}

\allowdisplaybreaks
\usepackage{multicol}
\usepackage{color}
\definecolor{rosso}{cmyk}{0,1,1,0.4}
\definecolor{rossos}{cmyk}{0,1,1,0.55}
\definecolor{rossoc}{cmyk}{0,1,1,0.2}
\definecolor{blu}{cmyk}{1,1,0,0.3}
\definecolor{blus}{cmyk}{1,1,0,0.6}
\definecolor{bluc}{cmyk}{1,1,0,0.1}
\definecolor{verde}{cmyk}{0.92,0,0.59,0.25}
\definecolor{verdec}{cmyk}{0.92,0,0.59,0.15}
\definecolor{verdes}{cmyk}{0.92,0,0.59,0.4}

\newcommand{\eq}[1]{~{\rm (\ref{eq:#1})}}

\newcommand{\GeV}{\,{\rm GeV}}
\newcommand{\TeV}{\,{\rm TeV}}

\def\circa#1{\,\raise.3ex\hbox{$#1$\kern-.75em\lower1ex\hbox{$\sim$}}\,}

\newcommand{\beq}{\begin{equation}}
\newcommand{\eeq}{\end{equation}}

\newcommand{\bea}{\begin{eqnarray}}
\newcommand{\eea}{\end{eqnarray}}
\newcommand{\be}{\begin{equation}}
\newcommand{\ee}{\end{equation}}
\font\tenrsfs=rsfs10 at 12pt
\font\sevenrsfs=rsfs7
\font\fiversfs=rsfs5
\newfam\rsfsfam
\textfont\rsfsfam=\tenrsfs
\scriptfont\rsfsfam=\sevenrsfs
\scriptscriptfont\rsfsfam=\fiversfs

\newsavebox\MBox

\renewenvironment{thebibliography}[1]
{\begin{multicols}{2}[\section*{\refname}]%
		\@mkboth{\MakeUppercase\refname}{\MakeUppercase\refname}%
		\list{\@biblabel{\@arabic\c@enumiv}}%
		{\settowidth\labelwidth{\@biblabel{#1}}%
			\leftmargin\labelwidth
			\advance\leftmargin\labelsep
			\@openbib@code
			\usecounter{enumiv}%
			\let\p@enumiv\@empty
			\renewcommand\theenumiv{\@arabic\c@enumiv}}%
		\sloppy
		\clubpenalty4000
		\@clubpenalty \clubpenalty
		\widowpenalty4000%
		\sfcode`\.\@m}
	{\def\@noitemerr
		{\@latex@warning{Empty `thebibliography' environment}}%
		\endlist\end{multicols}}

\newcommand{\SU}{\,{\rm SU}}

\newcommand{\U}{\,{\rm U}}
\renewcommand{\L}\Lag

\def\circa#1{\,\raise.3ex\hbox{$#1$\kern-.75em\lower1ex\hbox{$\sim$}}\,}
\makeatletter

\font\ital=cmu10

\def\hhref#1{\href{http://arxiv.org/abs/#1}{arXiv:#1}}
\usepackage{xstring}
\newcommand{\hhrefq}[1]{\IfSubStr{#1}{:}{\href{http://inspirehep.net/search?ln=en&ln=en&p=#1&of=hb&action_search=Search&sf=&so=d&rm=&rg=25&sc=0}{InSpire:#1}}{\hhref{#1}}}

\def\art{\@ifnextchar[{\eart}{\oart}}
\def\eart[#1]#2#3#4#5#6{{\rm #2}, {\em #3 \bf #4} {\rm (#6) #5} ({\em #1})}
\def\article{\@ifnextchar[{\earticle}{\oarticle}}
\def\oarticle#1#2#3#4#5#6{{\rm #1}, {\ital `#6'}, {\rm #2 #3 (#5) #4}}
\def\earticle[#1]#2#3#4#5#6#7{{\rm #2}, {\ital `#7'}, {\rm #3 #4 (#6) #5}  [\hhrefq{#1}]}
\def\hepart[#1]#2{{\rm #2, \sl#1}}
\def\heparticle[#1]#2#3{#2, {\ital `#3'} [\hhrefq{#1}]}
\newcommand{\doi}[1]{\href{http://dx.doi.org/#1}{[link]}}

\newcommand{\hhrefqq}[1]{\IfBeginWith{#1}{10.}{\href{https://doi.org/#1}{doi:#1}}{\hhrefq{#1}}}
\def\earticle[#1]#2#3#4#5#6#7{{\rm #2}, {\ital `#7'}, {\rm #3 #4 (#6) #5}  [\hhrefqq{#1}]}
\renewenvironment{thebibliography}[1]
{\begin{multicols}{2}[\section*{\refname}]%
		\@mkboth{\MakeUppercase\refname}{\MakeUppercase\refname}%
		\list{\@biblabel{\@arabic\c@enumiv}}%
		{\settowidth\labelwidth{\@biblabel{#1}}%
			\leftmargin\labelwidth
			\advance\leftmargin\labelsep
			\@openbib@code
			\usecounter{enumiv}%
			\let\p@enumiv\@empty
			\renewcommand\theenumiv{\@arabic\c@enumiv}}%
		\sloppy
		\clubpenalty4000
		\@clubpenalty \clubpenalty
		\widowpenalty4000%
		\sfcode`\.\@m}
	{\def\@noitemerr
		{\@latex@warning{Empty `thebibliography' environment}}%
		\endlist\end{multicols}}

\newcounter{alphaequation}[equation]
\def\thealphaequation{\theequation\hbox to
	0.6em{\hfil\alph{alphaequation}\hfil}}
\def\eqnsystem#1{
	\def\@eqnnum{{\rm (\thealphaequation)}}
	\def\@@eqncr{\let\@tempa\relax \ifcase\@eqcnt \def\@tempa{& & &} \or
		\def\@tempa{& &}\or \def\@tempa{&}\fi\@tempa
		\if@eqnsw\@eqnnum\refstepcounter{alphaequation}\fi
		\global\@eqnswtrue\global\@eqcnt=0\cr}
	\refstepcounter{equation} \let\@currentlabel\theequation \def\@tempb{#1}
	\ifx\@tempb\empty\else\label{#1}\fi
	\refstepcounter{alphaequation}
	\let\@currentlabel\thealphaequation
	\global\@eqnswtrue\global\@eqcnt=0 \tabskip\@centering\let\\=\@eqncr
	$$\halign to \displaywidth\bgroup \@eqnsel\hskip\@centering
	$\displaystyle\tabskip\z@{##}$&\global\@eqcnt\@ne
	\hskip2\arraycolsep\hfil${##}$\hfil& \global\@eqcnt\tw@\hskip2\arraycolsep
	$\displaystyle\tabskip\z@{##}$\hfil
	\tabskip\@centering&\llap{##}\tabskip\z@\cr}
\def\endeqnsystem{\@@eqncr\egroup$$\global\@ignoretrue} \makeatother

\definecolor{Gray}{gray}{0.95}

\def\bal#1\eal{\begin{align}#1\end{align}}

\begin{document}
\begin{center}  
{\huge\bf\color{rossos} Interpreting electroweak precision data} \\[1ex]
{\huge\bf\color{rossos} including the $W$-mass CDF anomaly}\\[3ex]
{\bf\large Alessandro Strumia}\\[2ex]
{\it Dipartimento di Fisica, Universit\`a di Pisa, Italia}\\[3ex]
{\large\bf Abstract}\begin{quote}
We perform a global fit of electroweak data, finding that the anomaly in the $W$ mass claimed by the CDF collaboration
can be reproduced as a universal new-physics
correction to the $T$ parameter or $| H^\dagger D_\mu H|^2$ operator. 
Contributions at tree-level from multi-TeV new physics can fit the anomaly compatibly with collider bounds:
we explore which scalar vacuum expectation values
(such as a triplet with zero hypercharge), 
$Z'$ vectors (such as a $Z'$ coupled to the Higgs only), little-Higgs models or higher-dimensional geometries 
provide good global fits.
On the other hand, new physics that contributes at loop-level 
must be around the weak scale to fit the anomaly.
Thereby it generically conflicts with collider bounds,
that can be bypassed assuming special kinematics like quasi-degenerate particles that decay into Dark Matter
(such as an inert Higgs doublet or appropriate supersymmetric particles).
\end{quote}
\end{center}
\tableofcontents

\section{Introduction}
\begin{figure}[t]
\begin{center}
$$\includegraphics[width=0.75\textwidth]{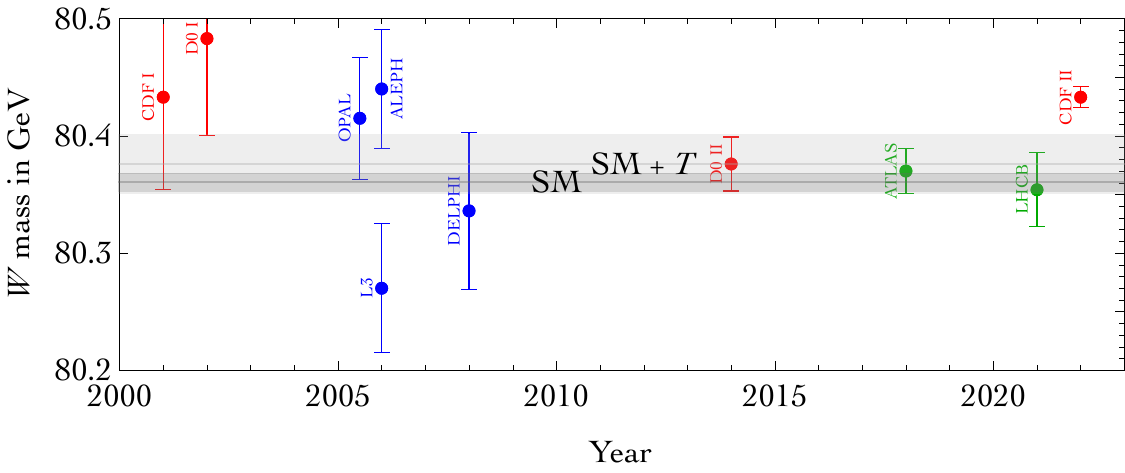}$$
\caption{Measurements of the $W$-boson mass compared to the $\pm 1\sigma$ band predicted by the
SM  (darker thinner band)
and by the SM extended allowing for a free $T$ parameter (lighter wider band and eq.\eq{MWSMT}).
\label{fig:MWexp}}
\end{center}
\end{figure}

The CDF collaboration announced an updated more precise measurement of the $W$-boson mass~\cite{CDF:2022hxs},
\beq M_W = (80.433 \pm 0.0064_{\rm stat}\pm 0.0069_{\rm syst})\GeV,\eeq
that shows a significant $7\sigma$ disagreement with the Standard Model (SM) prediction,
$M_W =(80.357\pm 0.006)\GeV$, but also
disagrees with the previous global combination of data from LEP, CDF, D0 and ATLAS,
$M_W=80.379\pm 0.012$~GeV~\cite{ParticleDataGroup:2020ssz}.
The situations is illustrated in fig.\fig{MWexp}.
Waiting for a future CMS measurement of $M_W$, we assume
that the new CDF $M_W$ measurement is correct,
include it in a global fit of electroweak data, 
and explore which new physics is suggested by this anomaly,
compatibly with all other bounds.

\medskip

Section~\ref{heavy} shows how the anomaly can be reproduced by universal new physics.
Section~\ref{treeloop} shows that models where the new physics contributions arise at tree level
can easily satisfy collider bounds, while new physics contributions at loop level have generic problems
that can only be evaded in special situations.
Having clarified these main issues, section~\ref{Z'} shows how extra heavy $Z'$ vectors can fit the anomaly.
Finally, in section~\ref{LH} we show how the desired $Z'$ vectors (and other effects) are present in some
little-Higgs models proposed in the literature, finding that acceptable fits are possible.
We also comment on related extra-dimensional geometries.
Section~\ref{concl} contains our conclusions.

\section{Universal new physics}\label{heavy}
The Standard Model predicts the $W$ mass as
$M_W = M_Z \cos\theta_{\rm W}$ plus quantum corrections.
This means that the CDF anomaly could be directly due to  the $W$ mass, or
indirectly to the $Z$ mass, or to the weak angle $\theta_{\rm W}$, or to a new-physics  modification of the relation among them.
The $Z$ mass and the weak angle are measured very precisely. 
Loop corrections also depend on the top mass, on the Higgs mass and on
the strong coupling, that again are measured precisely enough. 
For example, to reproduce the CDF measurement within the SM, the top mass would need
to be about $11\GeV$ than its measured value.
The Higgs mass too is now measured very precisely.
Thereby the CDF $M_W$ anomaly needs new physics.

\smallskip

The key issue is weather some new physics can account for the CDF anomaly compatibly with all precision data
and with LHC measurements at  higher energy. 
The answer is yes, and the new physics that can fit the CDF anomaly is simple,
of a type known as {\em heavy universal new physics at leading order.
`Heavy' means that it can be described as effective operators;
`universal' means operators that only involve the weak gauge bosons and the Higgs;
`leading order' means non-renormalizable operators with lowest dimension 6.\footnote{Operators with dimension 5
cannot be used, because the SM only allows for those that break lepton number, leading to neutrino masses.}
%\AS{At one loop $\widehat T \simeq 3 g_2^2 M_t^2/64\pi^2 M_W^2$ (dependence on $M_t$ slig reduced by higher loops)
%so the anomaly is like a $11\GeV$ shift in the top mass. The Higgs mass is measured more precisely and is less relevant.}
%as $T$ but suppressed, unless in special models. The only advantage is that it affects $M_W$ only.
This kind of new physics} can be parameterized adding to the SM Lagrangian the four effective  {\rm SU(2)$_L$}-invariant  dimension-6 operators 
listed in table~\ref{tab:STWY}
\begin{equation}
\mathscr{L}_{\rm eff}=\mathscr{L}_{\rm SM}+ 
\frac{1}{v^2}\bigg[  c_{WB} 
 {\cal O}_{WB}+
 c_{H} {\cal O}_{H}+c_{WW} {\cal O}_{WW}+
 c_{BB} {\cal O}_{BB}\bigg]\, .
\label{eq:NRO}
\end{equation}
Here $v\approx 174 \GeV$ is the Higgs vacuum expectation value, $\med{H}=(0,v)$,
and the coefficients can be conveniently written in terms of corrections to effective SM vector boson propagators as
\begin{equation}
\widehat{S}=2\frac{\cW}{\sW}c_{WB}\ ,\qquad
\widehat{T}=-c_{H}\ , \qquad
W = -g^2 c_{WW}\ ,\qquad 
Y = - g^2c_{BB}
\end{equation}
where the $\widehat S$, $\widehat T$ , $W$ and $Y$ coefficients are defined in table~\ref{tab:STWY}.
The SM corresponds to $\widehat S=\widehat T =W=Y = 0$.
Our $\widehat{S}$, $\widehat{T}$ are related to the
  usual $S,T$ parameters~\cite{Peskin:1990zt} as    $S= 4s_{\rm W}^2  \widehat{S}/\alpha\approx 119\, \widehat{S}$ and
  $T= \widehat{T}/\alpha\approx 129\, \widehat{T}$.
  The often-considered $U$ parameter
  corresponds to a dimension 8 effective operator analogous to ${\cal O}_H$ but with two extra derivatives.
On the other hand, the $W$ and $Y$ parameters need to be included to describe universal dimension-6 operators~\cite{hep-ph/0405040}.

\subsection{Summary of data} 
Our data-set includes all traditional precision electroweak data, including  the recent improved computation of the bottom forward/backward asymmetry~\cite{2003.13941}, and measurements of the Higgs and top-quark masses
\beq M_h = (125.1\pm 0.2)\GeV~\hbox{\cite{ParticleDataGroup:2020ssz}}\qquad M_t =(172.4\pm 0.3)\GeV~\hbox{\cite{ParticleDataGroup:2020ssz,CMStop22}}.\eeq
The top mass combines the  latest Particle Data Group average~\cite{ParticleDataGroup:2020ssz}
with the new CMS result~\cite{CMStop22}.
%$M_t = (171.77\pm 0.38)\GeV$
%$M_t = (172.76\pm 0.30)\GeV$) 
We also include LEP2 data that constrain $W,Y$ at $10^{-3}$ level.
More sensitive probes to $W,Y$ arise from recent LHC data, that however have not yet been analysed in a systematic way.
We thereby include LHC data in the following approximated way.
We recall the identity ${\cal O}_{BB} = J_B^2$ and ${\cal O}_{WW} = J_W^2$
where $J \sim g [ i H^\dagger D_\mu H + \sum_{f} \bar f \gamma_\mu f ]$ are the Higgs plus fermion
$f=\{q,\ell\}$ currents coupled to $\U(1)_Y$ and $\SU(2)_L$ vectors, respectively~\cite{hep-ph/9906266}.
So $W,Y$ are equivalent to specific combinations of current-current dimension-6 operators.
Generically speaking, thanks to its higher energy LHC is now significantly more sensitive then LEP to operators of the form
$(\bar q \gamma_\mu q)(\bar \ell \gamma_\mu \ell)$, that manifest as $e^+ e^-\to jj$ at LEP and as $pp\to \ell \bar\ell$ at LHC.
Here $q$ denotes a generic quark and $\ell$ a generic lepton, with unspecified chiralities.
Furthermore LHC now starts competing with LEP on
operators of the form $(\bar q \gamma_\mu q)(H^\dagger D_\mu H)$, that manifest at LEP as
modifications of $Z \bar qq$ couplings, and at LHC as $pp \to VV$ where $V=\{W,Z,h\}$.
The LHC sensitivity is reduced by the reconstruction efficiency of $V$ in current analyses~\cite{1712.01310} (see~\cite{2012.02779} for a recent fit).
Overall, quark/lepton operators at LHC now provide the dominant sensitivity to $W,Y$, with the LHC result
\beq W = (-0.12\pm 0.06)\,10^{-3},\qquad |Y| \circa{<} 0.2 \,10^{-3} \hbox{ at 95\% C.L}.\label{eq:LHCWY}
\eeq
 The $W$ measurement comes from
$101/{\rm fb}$ of CMS data about $pp\to \ell \slashed{E}_T$ at $\sqrt{s}=13\TeV$~\cite{2202.06075}.
The bound on $Y$ is estimated by recasting the bounds on $Z'$ vectors
from $36/{\rm fb}$ of ATLAS $pp\to \ell^+\ell^-$ data at $\sqrt{s}=13\TeV$~\cite{1707.02424}.
We cannot extract the central value of $Y$, small and negligible.
Consistently with the sensitivity estimates of~\cite{1609.08157,2008.12978,2103.10532}
these LHC results significantly improve over LEP2, affecting our subsequent discussion.
Bounds from higher-energy LHC scatterings on $\widehat S$, $\widehat T$ are instead negligible.
Let us, for example, discuss the ${\cal O}_{H}=|H^\dagger D_\mu H|^2 $ operator that will allow to fit the CDF $M_W$ anomaly.
Picking its $H \to (0,v)$ and $D_\mu \to i A_\mu^a T^a$ part gives the $\widehat T$ parameter,
while picking $D_\mu\to \partial_\mu$ together with the fields in $H$ gives
energy-enhanced $2\to 2$ scatterings involving $h$ and the longitudinal components of the $W$ and $Z$ bosons,
such as $hh\to hh$ or $W_L W_L\to W_L W_L$.
LHC has limited sensitivity to these processes because of the suppression needed to get $hh$ or $WW$ collisions out of $pp$ collisions.
The other intermediate terms in ${\cal O}_{H}$ gives small corrections to Higgs decays of relative order $\widehat T$.

\medskip

% Y about twice more uncertain than W

\smallskip
 
We next include in our fit the new CDF $W$-mass measurement.
No univocal procedure dictates how to deal with the experimental $\approx 4\sigma$ inconsistency among different $M_W$ measurements. 
Prescriptions that artificially increase the uncertainty of the global average 
when combining seemingly incompatible measurements are justified under the assumption that some measurement is wrong.
If instead the discrepancy is due to unlikely fluctuations,
the weighted average gives the correct statistical implication of the data.
For the sake of simplicity we follow the majority of the literature: 
we only include the CDF $M_W$ result in order to explore its implications.
%by performing the weighted average of all $W$-mass measurements.
%While this can be questioned in view of the $4\sigma$ discrepancy between CDF and other measurements, 
%we have no reason to discard other measurements. 
The above choice has minor practical relevance: the weighted average would be
dominated by CDF thanks to its claimed smaller uncertainty.
The large ($7\sigma$) statistical significance of the CDF anomaly 
makes details of the fitting procedure less important than the key binary issue:
can the CDF anomaly be fitted by adding new physics?
In the next section we confirm that the answer is yes.

% fig. 1 of 1312.2928

\begin{figure}[p]
\begin{center}
$$\includegraphics[width=0.45\textwidth]{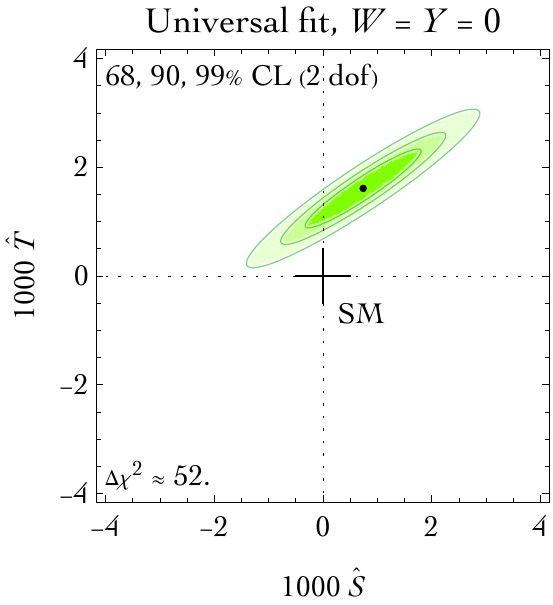}
%\qquad\includegraphics[width=0.45\textwidth]{figs/STwithMWanyST}$$
\includegraphics[width=0.45\textwidth]{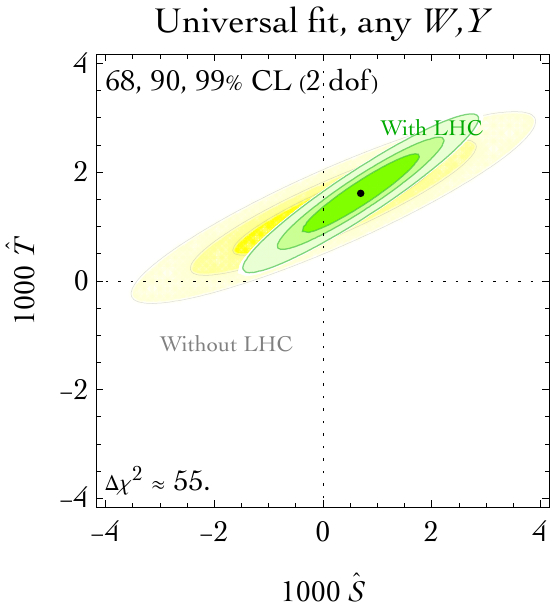}$$
\caption{The green regions are favoured by our global fit of the universal $\widehat S, \widehat T, W, Y$ electroweak parameters.
The yellow regions shwow the analogous results computed without including the new LHC bounds on $W,Y$.
\label{fig:ST}}
%\end{center}
%%\end{figure}
%%\begin{figure}[t]
%\begin{center}
$$\includegraphics[width=0.45\textwidth]{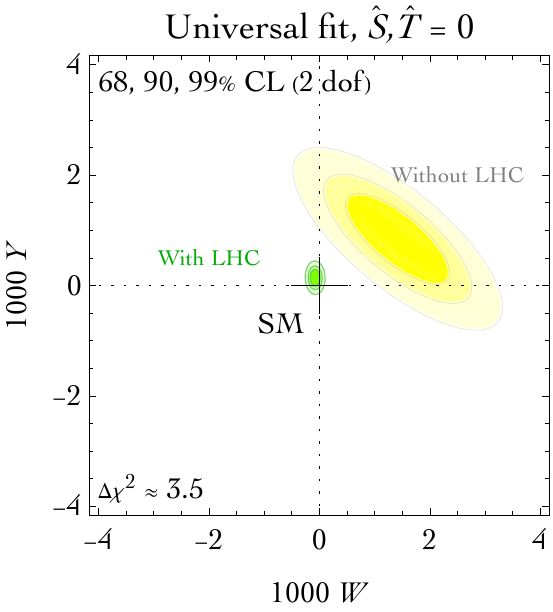}\qquad
\includegraphics[width=0.45\textwidth]{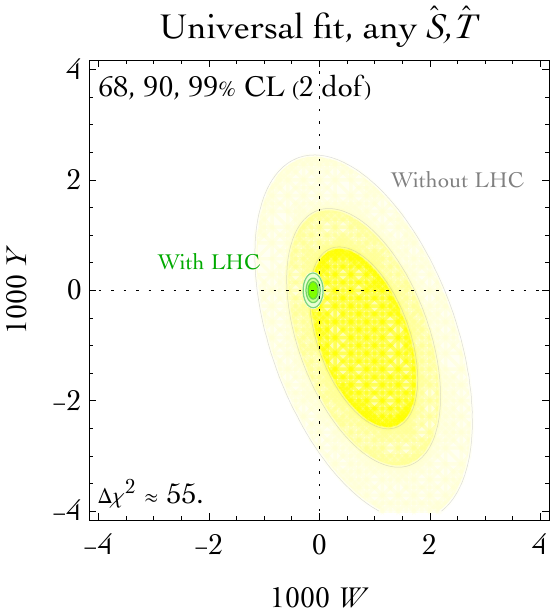}$$
\caption{Global fit of the universal $W$ and $Y$ electroweak parameters. Regions as in fig.\fig{ST}.
\label{fig:WY}}
\end{center}
\end{figure}

\begin{table}[t]
$$
\begin{array}{rclrlcc}
\multicolumn{3}{c}{\hbox{Dimension-less form factors}}&
\multicolumn{2}{c}{\hbox{operators}}\\ \hline
(g'/g){\color{blus}\widehat{S}} &=& \Pi'_{W_3 B}(0) & {\cal O}_{WB}~=&(H^\dagger \tau^a H) W^a_{\mu\nu} B_{\mu\nu} \!\!\!\\[1mm]
M_W^2{\color{blus} \widehat{T} }&=& \Pi_{W_3 W_3}(0)-\Pi_{W^+W^-}(0)\!\!\!
& {\cal O}_H~=&|H^\dagger D_\mu H|^2\\[1mm]
2M_W^{-2}{\color{blus} Y} &=&\Pi''_{BB}(0) &{\cal O}_{BB}~=&(\partial_\rho B_{\mu\nu})^2/2 = J_B^2\\[1mm]
2 M_W^{-2}{\color{blus} W} &=& \Pi''_{W_3 W_3}(0) & {\cal O}_{WW} ~=&(D_\rho W^a_{\mu\nu})^2/2 = J_W^2 \\[1mm]
\end{array}$$
  \caption{\label{tab:STWY}   The first column defines the dimension-less form factors.
The second column defines the {\rm SU(2)$_L$}-invariant universal dimension-6 
operators,
  which contribute to the form-factors on the same row.
  We use canonically normalized fields and inverse propagators $\Pi(k^2)= k^2 - M^2 + \cdots$
  and denote as $g, g'$ the $\SU(2)_L\otimes{\rm U}(1)_Y$ SM gauge couplings.}
\end{table}

\subsection{Global fit: SM plus free $\widehat{T}$ parameter}
In view of the CDF anomalous $M_W$ measurement,
the global fit now favours at high confidence level a new physics effect.
It could be due to 
$ \widehat T \approx (1.13\pm 0.16) \,10^{-3}$  alone,
corresponding to a new-physics effective operator $\Lag \approx \Lag_{\rm SM}-  |H^\dagger D_\mu H|^2/(5\TeV)^2$.
We considered only CDF as $W$-mass measurement in the global fit. The resulting $\chi^2_{\rm CDF~only}$
can be compared with $\chi^2_{\rm no~CDF}$, obtained including in the global fit
the weighted average of all $W$-mass measurements but CDF.
The result is:
\beq [\chi^2_{\rm CDF~only} -  \chi^2_{\rm no~CDF}]_{\rm SM}\approx 7.2^2,\qquad
 [\chi^2_{\rm CDF~only} -  \chi^2_{\rm no~CDF}]_{{\rm SM}+\widehat{T}} \approx 2.2^2.
\eeq
The large difference in the $\chi^2$ of the SM fits just means that CDF finds a large $\sim 7\sigma$ deviation from the SM.
The small difference in the $\chi^2$ when the SM is extended by allowing for a
free $\widehat{T}$ parameter 
%(and the mildly smaller difference in the more general fits where also $\widehat{S}$ and possibly $W,Y$ are free) 
means that this extended theory can reasonably account for the CDF anomaly.
This same conclusion is reached by excluding all $M_W$ measurements and
assuming as theory the SM plus a free $\widehat T$ parameter: the predicted $W$-mass is
\beq M_W = (80.376\pm 0.025)\GeV\qquad\hbox{(SM extended allowing for a free $\widehat{T}$ parameter)}
 \label{eq:MWSMT}\eeq 
%the weighted average of all measurements would give $ \widehat T \approx (0.88\pm 0.14) \,10^{-3}$.} 
plotted in fig.\fig{MWexp} as a thicker grey band.
This again shows that, mostly in view of the higher uncertainty in eq.\eq{MWSMT}, 
this SM extension is enough to accomodate the CDF anomaly at about $2\sigma$ level. 
The $M_W$ prediction of eq.\eq{MWSMT} has a larger but finite uncertainty because $\widehat T$ also affects the Fermi constant measured at zero momentum.\footnote{An interesting feature of the neglected $U$ parameter 
is that it only affects the $W,Z$ masses involved in the CDF anomaly.
Eq.\eq{MWSMT} shows that this feature is not needed to fit the $W$-mass anomaly: $\widehat T$ does a good enough job.
Furthermore, devising models where $U$ is significant is a difficult task, 
because $U$ has the same symmetry properties as $\widehat T$ so that
the higher dimensionality of $U$ implies that new physics with mass $m$ generates 
a suppressed $U/\widehat T \sim  \min(1,M_W^2/m^2)$.
Thereby $U$ is only relevant if $m\sim M_W$. New electroweak physics is this mass range is now mostly excluded by LHC data.}

\subsection{Global fit: SM plus free $\widehat{S},\widehat{T},W,Y$ parameters}
Next, returning to our global fit, and allowing also for a non-vanishing $\widehat S$ together with $\widehat T$, 
the fit in  fig.\fig{ST}a finds that a positive  $\widehat S$ comparable to $\widehat T$ is allowed by data: 
\beq \widehat T \approx (1.2\pm 0.5) \,10^{-3},\qquad \widehat S \approx (0.5 \pm 0.7) \,10^{-3}\eeq
%\beq \widehat T \approx (1.6\pm 0.5) \,10^{-3},\qquad \widehat S \approx (0.7 \pm 0.7) \,10^{-3}\eeq
with strong correlation $\rho \approx 0.94$.
The correlation and the best-fit values can be approximatively understood
by noticing that the anomaly $\delta M_W/M_W \approx 0.7\, \permille $ claimed by CDF
is so much statistically significant that reproducing it via the theoretical formula
$\delta M_W/M_W \simeq (c_{\rm W}^2 \widehat T/2  -s_{\rm W}^2 \widehat S) /(c_{\rm W}^2-s_{\rm W}^2)+\cdots$
dominates the global fit.
The confidence levels of fig.\fig{ST} are computed
knowing that $\chi^2 -\chi^2_{\rm best} $ follows a $\chi^2$ distribution with 2 degrees of freedom
in the common Gaussian limit where the Bayesian and the frequentist approaches to statistical inference 
become independent of their arbitrary assumptions.
Since the SM extension can fit the anomaly, the SM point $\widehat S=\widehat T=0$ is strongly disfavoured.

The difference $\Delta\chi^2 = \chi^2_{\rm SM}-\chi^2_{\rm best} $ quoted in fig.\fig{ST}a
or the pulls of the various observables show that universal new physics is enough to 
reproduce the observables (including $M_W$) obtaining an overall good quality of the fit, 
except of course for the internal inconsistency between different measurements of $M_W$, all included in our fit.
Fig.\fig{ST}a also shows that a poorer fit is obtained for vanishing $\widehat T=0$ and $\widehat S<0$. 
Roughly the same strong preference for $\widehat T>0$ and its correlation with $ \widehat S$ 
hold in a more general fit that marginalises over generic values of $W,Y$, as shown in fig.\fig{ST}b.
Fig.\fig{WY}a shows how
the strong new LHC bounds of eq.\eq{LHCWY} on $W,Y$ prevent the possibility of achieving a reasonable fit to the $M_W$ anomaly 
in the custodial-invariant limit $ \widehat T= \widehat S=0$.
Finally, fig.\fig{WY}b shows that the $M_W$ anomaly negligibly impacts the fit of $W,Y$.

\smallskip

The LHC bounds on $W,Y$ benefit from the large LHC energy, and are thereby
not applicable if the new-physics particles are so light that the effective field theory approximation breaks down at LHC.
In such a case, the yellow contours in fig.s\fig{ST},\fig{WY} apply, together with
extra LHC bounds on the production of the specific light particles.
Light new physics is needed if the $M_W$ anomaly is due to loop effects, as discussed in the next section.

%improving the $\chi^2$ with respect to the SM fit by the more modest amount shown in fig.\fig{WY}b.

%correspond precisely to 
%$$\Delta M_W = M_W/2 \alpha/(c^2 -s^2) ( S/2 + c^2 T ) $$

\section{New physics: at tree or loop level?}\label{treeloop}
Various particles with masses $m$ and couplings $g$
provide loop corrections to precision data. 
Since any $\SU(2)_L$ multiplet becomes quasi-degenerate in the  $m \gg v$ limit,
their effects are generically estimated as\footnote{This immediately follows
from the definitions in table~\ref{tab:STWY}
in terms of coefficients of dimension-6 operators.
The equivalent definition of the $\widehat{T}$ parameter in table~\ref{tab:STWY}
in terms of $W,W_3$ propagators at zero momentum leads to
$\widehat{T} \approx g_2^2 \Delta m^2/(4\pi M_W)^2$.
Eq.\eq{STWYest} is recovered taking into account that
the mass splitting among the components of the $\SU(2)_L$ multiplet in the loop
arises from couplings $g$ to the Higgs boson as $\Delta m\approx g^2 v^2/m$ in the limit $m\gg v$.} 
\beq \widehat{S},\widehat{T},W,Y \approx \frac{g^4 v^2}{(4\pi m)^2}  \label{eq:STWYest}.\eeq
The $M_W$ anomaly, $\widehat{T} \approx 10^{-3}$, is thereby reproduced for
$m/g^2 \approx 400\GeV$.
New physics in this mass range and significantly coupled to SM particles
is nowadays mostly excluded by LHC collider bounds,
altought dedicated searches are needed for models that only provide hidden signals.
For example special kinematics, such as decays into invisible quasi-degenerate particles,
tend to leave `holes' in exclusion bounds.

Let us consider the well known case of supersymmetric particles.
Their corrections to electro-weak parameters can be written analytically
in the limit of sparticle masses much heavier than the weak scale, $m\gg M_W$~\cite{hep-ph/0502095}.
This limit is nowadays relevant in view of collider bounds.
The supersymmetric correction to the $\widehat{T}$ parameter is~\cite{hep-ph/0502095}
\beq
\widehat{T} = \widehat T_{\rm sfermions}+ \widehat T_{\rm Higgses} + \widehat T_{\rm Higgsinos}^{\rm gauginos}\eeq
with
\begin{eqnsystem}{sys:TSUSY}
\widehat T_{\rm sfermions} & \simeq & 
\frac{\alpha_2}{16\pi}\frac{(M_t^2 + M_W^2\cos2\beta)^2}{m_{Q_3}^2 M_W^2}+
\frac{\alpha_2}{16\pi} M_W^2 \cos^22\beta\bigg( \frac{1}{m_L^2}+\frac{2}{m_Q^2}\bigg)\\
 \widehat T_{\rm Higgses}  & \simeq &
\frac{\alpha_2}{48\pi} \frac{M_W^2}{m_A^2}(1-\frac{M_Z^2}{M_W^2}\sin^2 2\beta)\\
\widehat T_{\rm Higgsinos}^{\rm gauginos}& \simeq &
\frac{\alpha_2 M_W^2}{48\pi M_2^2}\bigg[ \frac{7r-29+16r^2}{(r-1)^3}+
\frac{1+6r-6r^2}{(r-1)^4}6\ln r\bigg]\cos^2 2\beta
\end{eqnsystem}
where standard notations have been used,
$r = \mu^2/M_2^2$ and, just for simplicity, we assumed $M_1\gg|\mu|, M_2$.
The stop contribution, singled out in $\widehat T_{\rm sfermions}$,
is often dominant in view of its $g \approx y_t$ large couplings.
The stop (and sbottom) alone could fit the $M_W$ anomaly for $m_{Q_3}\sim 300\GeV$,
a range now mostly ruled even in some less visible kinematic configurations~\cite{1506.08616,1701.01954,2103.01290},
altought `holes' can remain in exclusion bounds.
Other sfermions can similarly contribute if huge trilinear couplings, not suppressed by the corresponding fermion masses, are assumed.

\smallskip

The Higgsino/gaugino system contributes to the $\widehat T$ parameter and provides an example of a system
where new particles can be hidden, as it contains a Dark Matter candidate and other charged states that 
can be quasi-degenerate to it, thereby decaying fast enough in a mostly-invisible channel.
More general similar examples can be built, for example restricting the models of~\cite{1111.2551} with a $\mathbb{Z}_2$ symmetry.
However, weakly interacting particles with weak-scale mass have a thermal relic
abundance smaller than the cosmological DM abundance and risk having a too large direct detection cross section
unless appropriate tunings are performed (see~\cite{hep-ph/0601041} for supersymmetric examples).

As another example, an `inert' Higgs doublet $H'=(h^\pm, (s+ia)/\sqrt{2})$
with components splitted by potential interactions 
$V = m^2 |H'|^2+\lambda_4 |H^* H'|^2 + \lambda_5 [(H^* H')^2+\hbox{h.c.}]/2 + \cdots$
with the SM Higgs $H$
contributes mostly to $\widehat T$ as~\cite{hep-ph/0603188}
\beq \widehat{T} = \frac{F(m_{h^\pm},m_a)+F(m_{h^\pm},m_s)-F(m_a,m_s)}{32\pi^2 v^2}
\stackrel{m\gg v}{\simeq} \frac{(m_{h^\pm}-m_a)(m_{h^\pm}-m_s)}{24\pi^2 v^2} \simeq
\frac{(\lambda_4^2-\lambda_5^2) v^2}{96\pi^2 m^2}
\eeq
where $F(m_1,m_2)=(m_1^2+m_2^2)/2- m_1^2 m_2^2\ln(m_1^2/m_2^2)/(m_1^2-m_2^2)$.
%while $\widehat S$ and $U$ are small.
This reproduces the $M_W$ anomaly for mildly large $\lambda_4 \approx m/v$ such that
$(m_{h^\pm}-m_a)(m_{h^\pm}-m_s)\approx M_W^2$,
implying weak-scale masses.
 %$\widehat{S},\widehat{T},U,W,Y$.
No dedicated LHC search established if an inert Higgs in this mass range
is still allowed. A recast of different  LHC searches~\cite{1812.07913} produced significant but partial bounds.
%We do not perform a global fit to precision data, as it needs to take into account the $U$ parameter too.

\medskip

Before moving from one loop to tree effects, let us mention an intermediate possibility:
log-enhanced one-loop effects.
The renormalisation group equations in the SM plus dimension 6 effective operators have been computed 
in a series of works culminated in~\cite{1310.4838,1312.2014,1312.2928,1505.03706},
finding that renormalisation from a few TeV scale down to the weak scale
induces specific non-vanishing mixings at few $\%$  level.
The operator ${\cal O}_H$ motivated by the $W$-mass anomaly
does not induce any operator that is significantly more constrained,
and can be induced by poorly constrained operators such as $(\partial_\mu |H|^2)^2$.
In turn, this operator can be mediated at tree level by a singlet scalar coupled only to $|H|^2$
(see~\cite{1203.0237} for a model in this sense) and thereby poorly constrained.

\medskip

We next consider new-physics effects at tree-level.
A variety of new particles can mediate tree-level corrections to $\widehat T $
 and to the other electroweak precision parameters.
 Let us consider extra scalars with a neutral component that acquires a vacuum expectation value.
 \begin{itemize}
\item A scalar triplet $T$ with hypercharge 0 
gets a $\U(1)_{\rm em}$-preserving vacuum expectation $v_T$ aligned to $H$ from a $A\,HH^\dagger T$ cubic coupling,
and contributes via a dimension-6 operator to $\widehat{T}$ only with the desired sign, $\widehat{T}=2v_T^2/v^2 > 0$,
so that the anomaly can be fitted for $v_T \approx 3\GeV$.
Its mass $M_T \approx (A v^2/v_T)^{1/2}$ is well above LHC bounds~\cite{2003.07867} if $A \sim M_T$.
This scalar was considered e.g.~in~\cite{Lynn:1990zk,1307.8134,2010.02597}.

\item A triplet with hypercharge 1 (coupled as $HHT^*$) 
contributes as  $\widehat{T}=-2v_T^2/v^2$.
This scalar appears in type II see-saw and in some little-Higgs models.

\item A similar situation is found for scalar 4-plets $Q$.
The quadruplet $Q$ with smaller hypercharge $Y=1/2$ (that gets a vacuum expectation $v_Q$ from a quartic coupling $\lambda\,HH^\dagger H Q^*$) 
contributes as $\widehat T = 6 v_Q^2/v^2>0$  and can fit the $M_W$ anomaly for $v_Q\approx 2\GeV$.
Its mass  $M_Q \approx (\lambda v^3/v_Q)^{1/2}$ is mildly above the LHC sensitivity if $\lambda\sim 1$.
Avoiding collider bounds is more difficult because the 4-plet mediates an effective dimension-8 operator.
Scalar quadruplets have been considered e.g.\ in~\cite{1704.07851,2005.00059}.

\item The quadruplet $Q$ with larger hypercharge $Y=3/2$ (quartic coupling $HH H Q^*$) contributes as
$\widehat T = -6 v_Q^2/v^2<0$.  
\end{itemize}
Contributions to $\widehat T$ for general representations are given in~\cite{1605.08267}.
%The $M_W$ anomaly favours a positive correction to  $\widehat{T}$.
 In the next section we focus on a specific plausible tree-level source of a positive $\widehat{T}$:
an extra heavy $Z'$ vector boson.
The sign of $\widehat{T}$ can be intuitively understood as follows:
the $Z/Z'$ mass mixing reduces the lighter $Z$ mass, following the general behaviour of eigenvalues.

  \section{Extra $Z'$ vector bosons}\label{Z'}  
A generic $Z'$ vector is conveniently  characterized by the following parameters:
its  gauge coupling $g_{Z'}$,
its mass $M_{Z'}$
 and the $Z'$-charges 
$Z_H$, $Z_L$, $Z_E$, $Z_Q$, $Z_U$, $Z_D$
of the Higgs doublet $H$ and of the SM fermion multiplets $L,E,Q,U,D$.
We assume a flavour-universal $Z'$.
Unless the $Z'$ couples to fermions universally or proportionally to the existing SM vectors, the $Z'$ does not give precision corrections of universal type,
and thereby a more complicated global fit is needed.
However, in practice, precision data about quarks are less precise than precision data about leptons, 
so that the $Z'$ quark charges $Z_Q$, $Z_U$, $Z_D$ less significantly affect global electroweak fits. 
 Then, the dependence on the more important  $Z_H$, $Z_L$, $Z_E$ parameters
 can be conveniently condensed in 
 a reduced set of approximatively universal corrections that include~\cite{hep-ph/0604111}
\beq\label{eq:TSZ'}
\widehat T =\frac{4 M_W^2 g_{Z'}^2}{g^2 M_{Z'}^2} (Z_E-Z_H+Z_L)^2, \qquad
\widehat S = \frac{2 M_W^2 g_{Z'}^2}{g^2 M_{Z'}^2} (Z_E-Z_H+Z_L)(Z_E+2Z_L + \frac{g^2}{g'^2} Z_E) 
 \eeq
where $g$ and $g'$ are the SM weak gauge couplings.
The extra similar expressions for $W,Y$ and for other coefficients can be found in eq.~(4.2) of~\cite{hep-ph/0604111}.
All extra effects apart from the correction to $\widehat T$ vanish if the $Z'$
couples only to the Higgs, i.e.\ $Z_H\neq 0 $ and $Z_{L,E,Q,U,D}=0$.
Notice also that $\widehat{S}=\widehat{T}=0$ whenever $Z_E-Z_H+Z_L=0$, such that
the charged lepton Yukawa interactions $LEH^*$ are invariant under the extra ${\rm U}(1)_{Z'}$ symmetry,
avoiding the need of a model for their generation.

\begin{figure}[t]
\begin{center}
$$\includegraphics[width=0.3\textwidth]{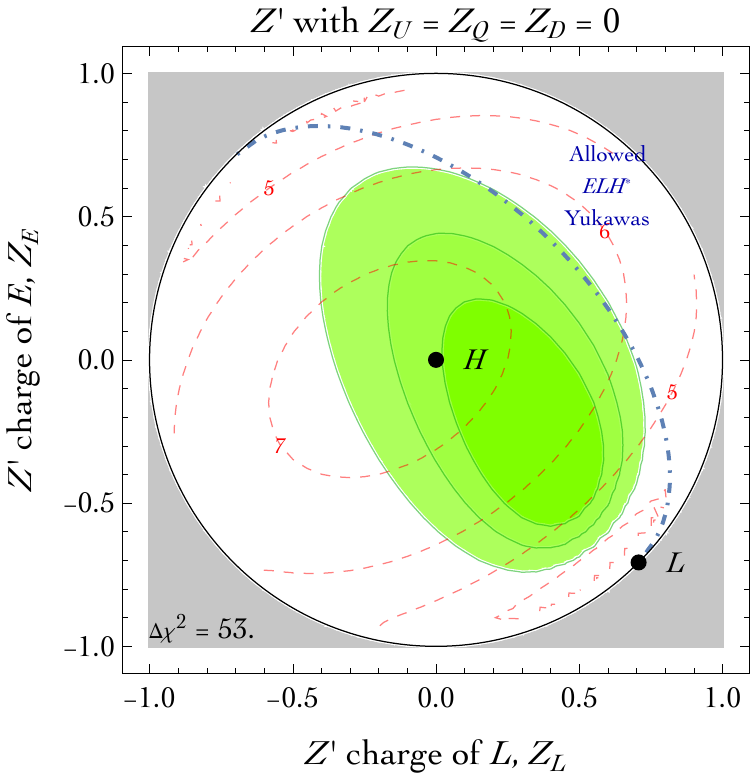}\qquad
\includegraphics[width=0.3\textwidth]{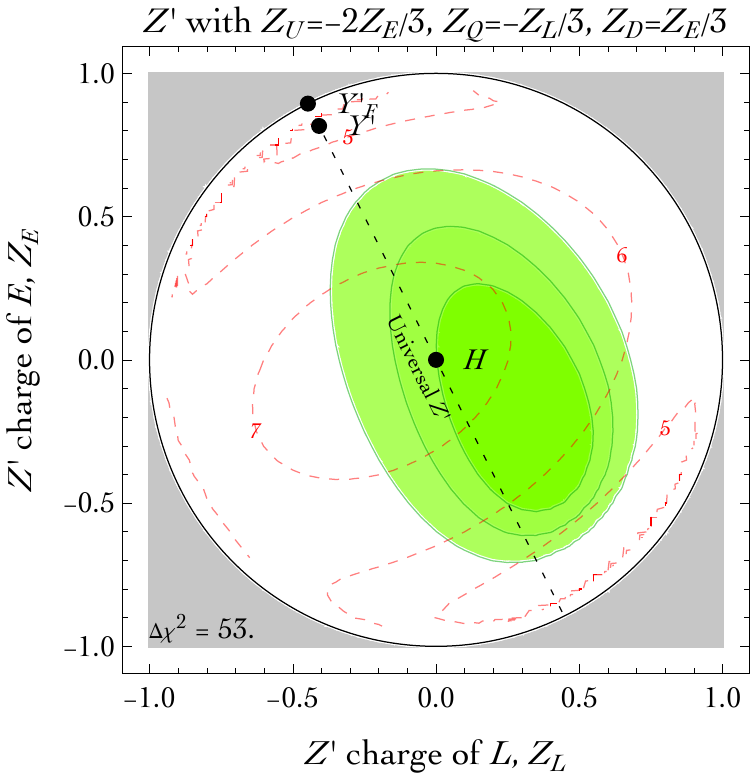}\qquad
\includegraphics[width=0.3\textwidth]{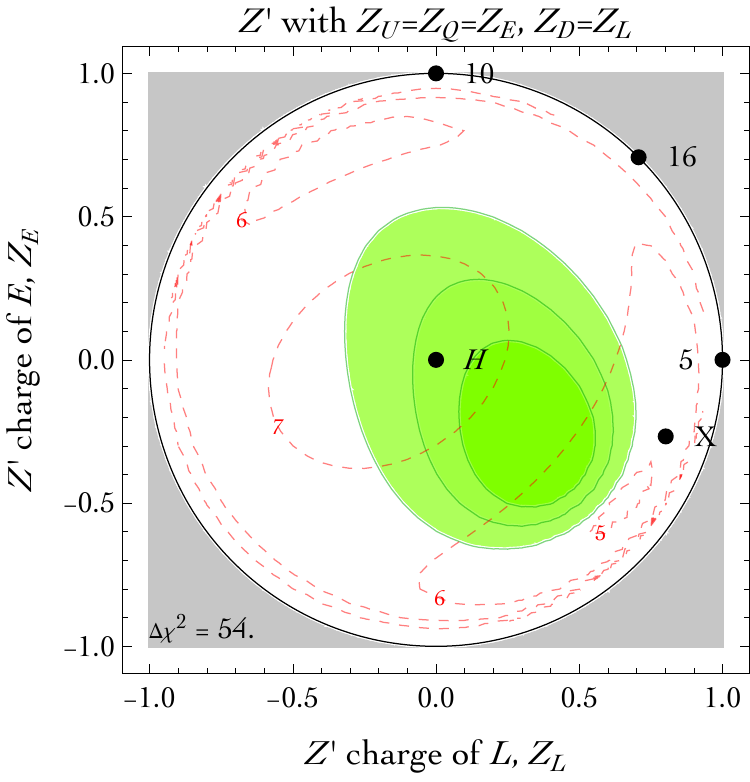}$$
\caption{\label{fig:Z'} Extra $Z'$ that fit electroweak data, including the $M_W$ anomaly.
The $Z'$ effects dominantly depend on the $Z'$ charge
of the Higgs and the leptons: we here assume the normalization
$Z_L^2 + Z_E^2 + Z_H^2 = 1$ such that $Z_H=0$ at the boundary of the circles,
and $Z_H=1$ in their centers.
The three panels assume different sets of quark charges, and show similar results.
The green contours show the best-fit regions at $68\%$ and $90\%$ confidence levels.
The red dashed contours show the best-fit values of $M_{Z'}/g_{Z'}$ in $\TeV$. 
The dashed line in the middle panel corresponds to a universal $Z'$
 and the dot-dashed curve in the first panel
to a $Z'$ compatible with the SM $LEH^*$ lepton Yukawas. 
The dots show some $Z'$s listed in figure~\ref{tab:famousZprimes}.}
\end{center}
\end{figure}

Barring this cancellation, the correction to $\widehat T $ has the desired sign.
The physical motivation is that the mass mixing of an heavy $Z'$ vector with the $Z$ vector reduces the $Z$ mass
while not affecting the $W$ mass, that thereby becomes relatively heavier compared to the $Z$.
So the anomaly $\widehat T \approx 10^{-3}$ is reproduced for 
\beq M_{Z'}/g_{Z'} \approx 8\TeV|Z_E-Z_H+Z_L|.\label{eq:Z'fit}\eeq
This simple approximation breaks down when $Z_E-Z_H+Z_L$ is small, so we show results of a global fit.
Without loss of generality we can normalise the $g_{Z'}$ coupling such that
$Z_H^2 + Z_L^2 + Z_E^2=1$ and assume $Z_H \ge 0$. 
Then 
we can compute the global $\chi^2$ and the best-fit value of $M_{Z'}/g_{Z'}$
on a half-sphere surface as function of the lepton charges $Z_L$ and $Z_E$, with $Z_H = \sqrt{1-Z_L^2-Z_H^2}$.

\smallskip

\begin{figure}[t]
\parbox{0.49\textwidth}{
$$
  \begin{array}{cc|cccccc}
\hbox{U(1)} & Z_H & Z_L & Z_D & Z_U & Z_Q & Z_E  \\ \hline
H  & 1 & 0 & 0 & 0 & 0 & 0 \\ 
Y' & \frac{1}{2} & -\frac{1}{2} & \frac{1}{3} & -\frac{2}{3} & 
\frac{1}{6} & 1 \\ 
Y'_F  & 0 & -\frac{1}{2} & \frac{1}{3} & -\frac{2}{3} & 
\frac{1}{6} & 1 \\ 
B-L  & 0 & -1 & -\frac{1}{3} & -\frac{1}{3} & \frac{1}{3} & 1 \\ 
L & 0 & 1 & 0 & 0 & 0 & -1 \\ 
10  & 0 & 0 & 0 & 1 & 1 & 1 \\ 
5  & 0 & 1 & 1 & 0 & 0 & 0 \\ 
X  & \frac{2}{3} & 1 & 1 & -\frac{1}{3} & -\frac{1}{3} & -\frac{1}{3} \\ 
16  & 0 & 1 & 1 & 1 & 1 & 1 \\
T_{3R}  & -\frac12 & 0 & -\frac12 & \frac12 & 0 & -\frac12\\
\chi & 2 &3 &3&-1 &-1 & -1 &  \\[0.5ex]
%\hbox{SLH} & \hbox{no} &  \multicolumn{6}{c}{\hbox{Simplest little Higgs~\cite{hep-ph/0407143}}} \\ 
%\hbox{SU6} & \hbox{no} &  \multicolumn{6}{c}{\hbox{Super little Higgs~\cite{hep-ph/0510294}}} \\ 
 \end{array}$$
\caption{\label{tab:famousZprimes}    Charges of frequently studied $Z'$ Proportionality constants have sometimes been omitted.}}~~
\parbox{0.49\textwidth}{$$\includegraphics[width=0.365\textwidth]{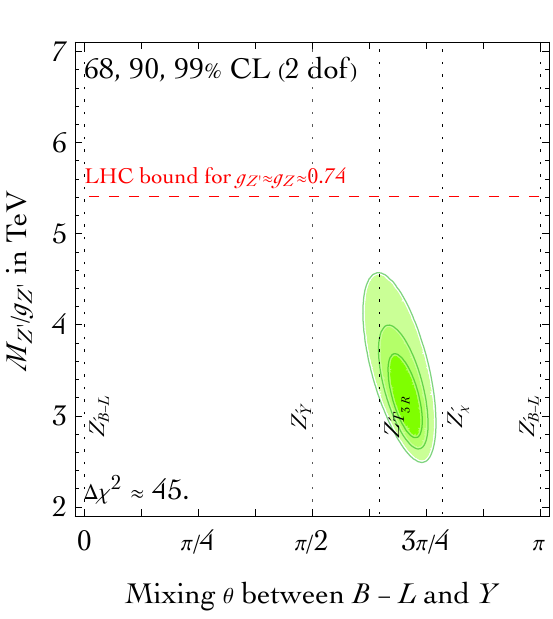}$$\caption{\label{fig:Zminimal}Minimal $Z'$ with charges
$Z_i=(B-L)_i\cos\theta + Y_i \sin\theta$
favoured by the global fit.}}
\end{figure}

Fig.\fig{Z'} shows the results of a global $Z'$ fit.
The three panels consider different assumptions for the less important quark $Z'$ charges:
zero in the left panel,
universal-like in the middle panel, and SU(5)-unified in the right panel
The panels exhibit similar results, confirming that $Z_H,Z_L$ and $Z_E$ are the most relevant parameters,
and that the universal approximation in eq.\eq{TSZ'} is accurate enough.
The left panel evades the strong LHC bounds on $(\bar q \gamma_\mu q)(\bar\ell \gamma_\mu \ell)$ operators,
that progressively become more relevant in the subsequent panels.
The dots in the plots highlight some commonly considered $Z'$ models listed in table~\ref{tab:famousZprimes}.
The best fit is provided by the universal $Z'$ denoted as $H$, as it corresponds to the Higgs only being charged under the $Z'$,
so that only $\widehat T>0$  gets corrected.

In view of the strong correlation between $\widehat T$ and $\widehat S$ found in section~\ref{heavy},
a wide variety of other $Z'$ models provide comparably good global fits to the $M_W$ anomaly.
Other effects, generically comparable to $\widehat T$, can be compatible with bounds.

\smallskip

From a theoretical point of view, it is interesting to consider a `minimal' $Z'$
with charges $Z_i=(B-L)_i\cos\theta + Y_i \sin\theta$
given by a linear combination of $B-L$ and hypercharge,
because this is the most generic flavour-universal and anomaly-free $Z'$ compatible with all SM Yukawa couplings~(see e.g.~\cite{0909.1320}).
While pure $B-L$ does not improve the SM fit, 
and a pure heavy hypercharge improves the fit only by mildly
(middle panel of fig.\fig{Z'}),
an acceptable fit is provided by an appropriate linear combination near $T_{3R}$,
see fig.\fig{Zminimal}. 
A lower $M_{Z'}/g_{Z'}$ is needed in view of the small value of $Z_H=\frac12\sin\theta$.
LHC restricts $M_{Z'}>4.1\TeV$ for $g_{Z'}=g_Z\approx 0.74$~\cite{1707.02424}.

\smallskip

We finally discuss if the $Z'$ vector bosons that fit the $M_W$ anomaly are compatible with collider bounds.
LEP2 $e^-e^+$ data are included in our global fit, and mostly constrain  4-lepton effective operators~\cite{hep-ph/0405040}.
LHC $pp$ data provide bounds~\cite{1707.02424} that
strongly depend on the quark $Z'$ charges that negligibly affect the electroweak global fit.
These collider bounds can be mostly avoided in models where $Z_{Q,U,D}$ vanish.
For generic values of  $Z_{Q,U,D}$ the qualitative situation is as follows:
collider bounds on $Z'$ production are more sensitive than precision data for
$Z'$ masses below about $4\TeV$~\cite{1707.02424}, while the limited LHC energy 
implies weaker sensitivity than precision data to heavier $Z'$.
As electroweak data only depend on the combination $M_{Z'}/g_{Z'}$,  
the $M_W$ anomaly can be fitted compatibly with collider bounds
for large enough $M_{Z'}$ masses corresponding to perturbative couplings $g_{Z'}\circa{>}0.5$
in view of eq.\eq{Z'fit}.
In this limit LHC sets bounds on
 $(\bar q\gamma_\mu q)(\bar \ell\gamma_\mu \ell)/\Lambda^2$ effective operators
 at the $\Lambda\circa{>}10\TeV$ level~\cite{1707.02424}.
These bounds, included in our fit (and used to estimate the bound on $Y$ in eq.\eq{LHCWY}), 
imply order one bounds on products of the quark and lepton $Z_{E,L,Q,U,D}$ 
charges for $Z'$ that fit the $M_W$ anomaly.
These bounds disfavour various motivated $Z'$, that have too large fermion charges.
Our fit does not include LHC data on $(\bar q\gamma_\mu q)(H^\dagger D_\mu H)$ operators,
that now provide bounds comparable to the bounds from LEP.

\section{Little Higgs models}\label{LH}
The operator $|H ^\dagger D_\mu H|^2$ motivated by the $M_W$ anomaly
generically arises in models where
the Higgs bosons is affected by new physics.
A plethora of particles that provide tree-level corrections to precision data
are predicted by models that were motivated by Higgs mass naturalness,
such as technicolour (where the Higgs becomes a bound state),
extra-dimensional models that allow TeV-scale quantum gravity
(where the Higgs supposedly becomes some stringy-like object).

\smallskip

We here focus on little-Higgs models that tried to obtain a naturally light  Higgs as the pseudo-Goldstone boson
of a suitable complicated pattern of symmetry breaking.
Such models contribute to electro-weak precision data at tree-level
that thereby prevent them from reaching  their naturalness goal.
A simple way of computing corrections to precision data in such models was described in~\cite{hep-ph/0502096},
where it was also noticed that many models are of `universal' type, allowing a unified systematic analysis.
In view of the $M_W$ anomaly, we reconsider those models that contribute to the $\widehat T$ parameter.
We focus on tree-level contributions due to $Z'$ and other heavy vectors, ignoring loop effects.

\subsection{The SU(5)/SO(5) `littlest' Higgs models}\label{sec:LittlestHiggs}
\begin{figure}[t]
\begin{center}
$$\includegraphics[width=0.43\textwidth]{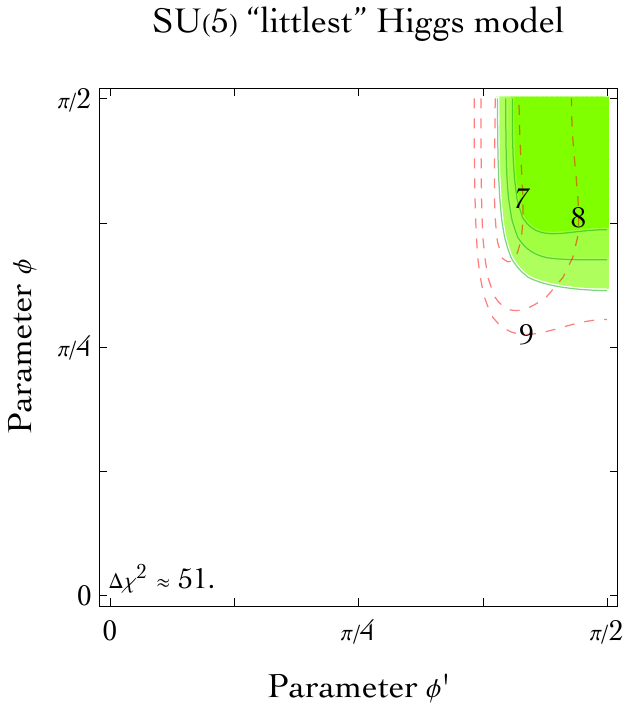}\qquad\includegraphics[width=0.43\textwidth]{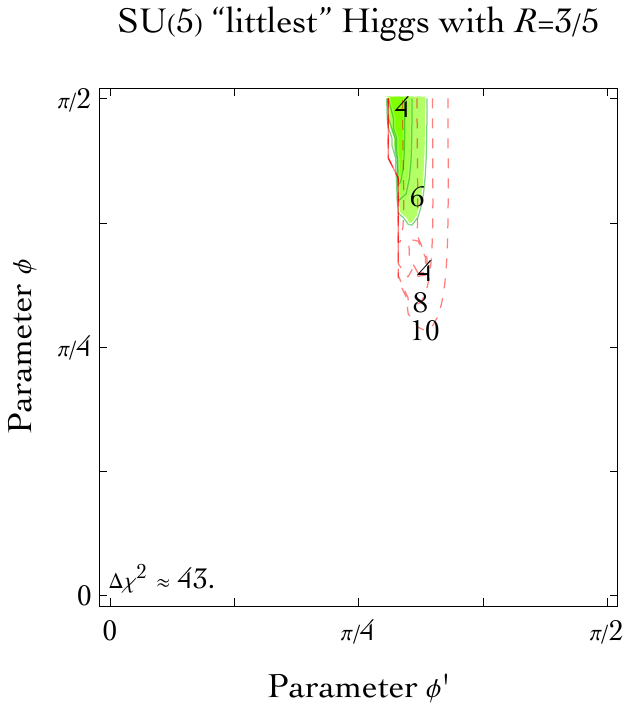}$$
\caption{Best-fit regions on the parameters of  $\SU(5)$ little-Higgs models.
The red dashed contours indicate the values of the scale $f$ in {\rm TeV}.
As described in the text, in each model the angles $\phi$ parameterize the gauge couplings of the
extra gauge groups, which become strongly coupled at $\phi,\phi'\to 0, \pi/2$.
\label{fig:LH}}
\end{center}
\end{figure}

The `littlest' Higgs model~\cite{hep-ph/0206021,Csaki:2002qg} assumes a SU(5) global symmetry broken to SO(5) at some scale $f$. 
The $\SU(2)_1\otimes \SU(2)_2 \otimes {\rm U}(1)_1 \otimes {\rm U}(1)_2$
subgroup of SU(5) is gauged, with gauge couplings
$g_1$, $g_2$, $g'_1$, $g'_2$ respectively.
The SM gauge couplings $g$ and $g'$ are obtained as
$1/g^2 = 1/g_1^2 + 1/g_2^2$ and $1/g^{\prime 2} = 1/g^{\prime 2}_1+1/g^{\prime 2}_2$.
The scale $f$ is normalized such that the extra heavy vector bosons have masses
\begin{equation}
  \label{eq:heavygb}
  M_{W'}^2 = \left( g_1^2 + g_2^2\right) \frac{f^2}{4}, \hspace{1cm} 
  M_{Y'}^2 = \left( g'^2_1 + g'^2_2 \right) \frac{f^2}{20}.
\end{equation}
Matter fermions are assumed to be charged  under $\SU(2)_1 \otimes {\rm U}(1)_1$ only.  
The model has three free parameters, which can be chosen to be $f$ and two angles $\phi$ and $\phi'$ defined as
\begin{equation}
 \sin \phi = g/g_1 \hspace{0.7cm}
  \cos \phi = g/g_2 \hspace{0.7cm}  \sin \phi' = g'/g'_1 \hspace{0.7cm}
  \cos \phi' = g'/g'_2.
\end{equation}
The universal corrections to precision data are~\cite{hep-ph/0502096}
\beq \label{eq:stwylittlest} \begin{array}{ll}\displaystyle
  \widehat S = \frac{2 M_W^2}{g^2 f^2} \bigg[\cos^2 \phi + 5\frac{\cW^2}{\sW^2} \; \cos^2
  \phi'\bigg], \qquad&\displaystyle
  W = \frac{4 M_W^2}{g^2 f^2} \cos^4 \phi,\\ \displaystyle
  \widehat T = \frac{5 M_W^2}{g^2 f^2}\ ,&
 Y = \displaystyle\frac{20 M_W^2}{g'^2 f^2} \cos^4 \phi'.
\end{array}\eeq
We omit a possible extra negative contribution to $  \widehat T$ from Higgs triplets with $Y=1$, present in this model.
Thereby the $M_W$ anomaly $  \widehat T \approx 10^{-3}$ can be fitted for $f\approx 9 \TeV$.
Fig.\fig{LH}a shows the best fit value of $f$ as function of $\phi$ and $\phi'$.  
The plots shows that a good global fit is obtained for small  
$\cos^2\phi$ and $\cos^2\phi'$, corresponding to larger $g_2$ and $g'_2$ couplings
and thereby to suppressed $W,Y$.
In this limit the $Y'$ vector can be heavy enough to be compatible with LHC bounds.

\bigskip

This `littlest Higgs' model can be modified by 
assigning charge $Y R$ under U(1)$_1$ and $Y (1-R)$ under U(1)$_2$ to the fermions.
Here $Y$ is the SM hypercharge and $0\le R\le 1$~\cite{Csaki:2002qg}.
The previous model corresponds to $R=1$.
Gauge interactions are not anomalous and are compatible with the needed SM
Yukawa couplings also for $R=3/5$~\cite{Csaki:2002qg}. 
The corrections to precision data become
% \SU(5)& 32211 & \displaystyle
% \frac{2 M_W^2}{g^2 f^2} \bigg[\cos^2 \phi + 5\frac{\cW^2}{\sW^2}(2R-1) \; (R-\sin^2
%   \phi')\bigg] &\displaystyle
%   \frac{5 M_W^2}{g^2 f^2} (2R-1)^2 +\widehat{T}_{\rm triplet}&\displaystyle
%    \frac{4 M_W^2}{g^2 f^2} \cos^4 \phi &
%    \displaystyle\frac{20 M_W^2}{g'^2 f^2} (R-\sin^2 \phi')^2\\[-3mm] \displaystyle
% R=3/5 &  &  & & \\[2mm] \displaystyle
\beq \label{eq:stwylittlest2} \begin{array}{ll}\displaystyle
  \widehat S = \frac{2 M_W^2}{g^2 f^2} \bigg[\cos^2 \phi + 5\frac{\cW^2}{\sW^2}(2R-1) \; (R-\sin^2\phi')\bigg],
 \qquad~&\displaystyle
  W = \frac{4 M_W^2}{g^2 f^2} \cos^4 \phi,\\ \displaystyle
  \widehat T = \frac{5 M_W^2}{g^2 f^2} (1-2R)^2\ ,& \displaystyle
 Y =\frac{20 M_W^2}{g'^2 f^2} (R-\sin^2 \phi')^2.
\end{array}\eeq
Fig.\fig{LH}b shows the global fit for $R=3/5$: LHC bounds on $W,Y$ are avoided for small $\cos\phi$ and $\sin^2\phi'\approx R$.
In view of the relatively smaller correction to $  \widehat T$, this modified model provides a less good fit to the $M_W$ anomaly.

%\bigskip

%The `littlest Higgs' model can also be modified by gauging only $\SU(2)_1\otimes \SU(2)_2 \otimes {\rm U}(1)_Y$~\cite{Perelstein2}. 
%In this way $\widehat{T}=0$ but one has a quadratic divergence to the Higgs mass associated to the hypercharge coupling $g'$. This model has been already analyzed in terms of $\widehat S,\widehat T,W,Y$ in~\cite{Barbieri:2004qk} and we report here the results
%\beq
%  \widehat S  = \frac{2 M_W^2}{g^2 f^2} \cos^2 \phi, \qquad
%    W  = \frac{4 M_W^2 }{g^2 f^2} \cos^4 \phi,\qquad
%  \widehat T  =Y  = 0.\eeq
%  Allowing for a generic $\widehat T$
%the $99\%$ C.L.\ constraint on $f$ is well approximated by
%$f >\max (6.5\cos^2\phi,3.7\cos\phi)\TeV$.
%In the limit of small $\cos\phi$ (which corresponds to large $g_2$)
%the constraint on $M_{W'}$ approaches a constant.
%$M_{W'}\circa{>}1.2\TeV$

\begin{figure}[t]
\begin{center}
$$\includegraphics[width=0.43\textwidth]{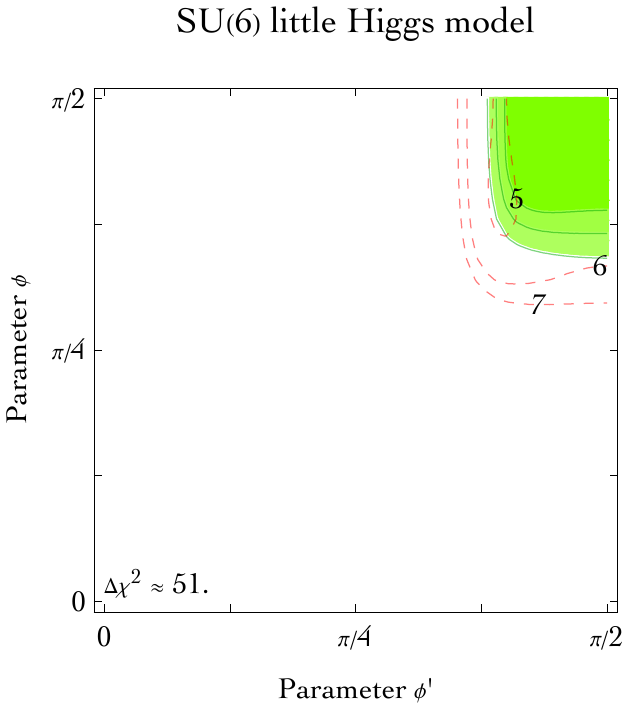}\qquad \includegraphics[width=0.43\textwidth]{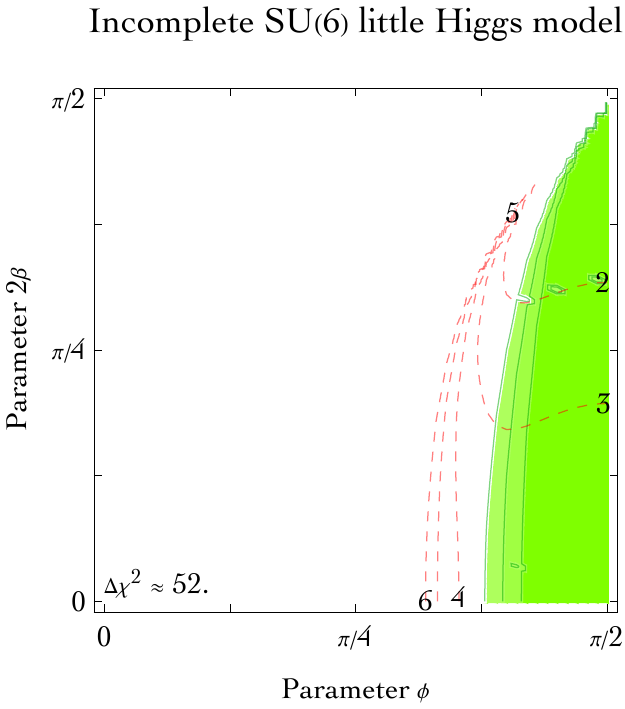}$$
\caption{As in fig.\fig{LH}, for $\SU(6)$ little-Higgs models.
\label{fig:LH6}}
\end{center}
\end{figure}

  \subsection{The SU(6)/Sp(6) models} \label{sec:SU6}
 This model~\cite{Low:2002ws} is based on a global symmetry 
 $\SU(6)$ broken to Sp$(6)$ at a scale $f$. 
The gauge group is $\SU(2)_1 \otimes \SU(2)_2 \otimes {\rm U}(1)_1 \otimes {\rm U}(1)_2$, 
with gauge couplings $g_1,g_2,g'_1,g'_2$,
broken to the diagonal $\SU(2)_L \otimes {\rm U}(1)_Y$ at the scale $f$. 
Following the notations of~\cite{Gregoire:2003kr}, the heavy gauge bosons have mass
\begin{equation}
  M_{W'}^2 = \left( g_1^2 + g_2^2\right) \frac{f^2}{4}, \hspace{1cm} 
  M_{Y'}^2= \left( g'^2_1 + g'^2_2\right) \frac{f^2}{8}
\end{equation}
and the SM gauge couplings are $1/g^2=1/g_1^2+1/g_2^2$ and $1/g'^2=1/g_1'^2+1/g_2'^2$.
This model contains no Higgs triplets. 
If the fermions are charged under $\SU(2)_1 \otimes {\rm U}(1)_1$ one gets:
\beq\begin{array}{ll}\displaystyle
   \widehat S = \frac{2 M_W^2}{g^2 f^2} \bigg[\cos^2 \phi +2  \frac{\cW^2}{\sW^2}
   \cos^2\phi' \bigg],\label{eq:S1}  \qquad & \displaystyle
  W  = \frac{4 M_W^2}{g^2 f^2} \cos^4 \phi, \\  \displaystyle
   \widehat T  = \frac{M_W^2}{2 g^2f^2}  (5+ \cos 4 \beta),&
   \displaystyle
   Y  = \frac{8 M_W^2}{g'^2 f^2} \cos^4 \phi'
  \end{array}\eeq
 where 
 \begin{align}
   & \cos \phi = g/g_1 \hspace{1cm} \sin \phi = g/g_2 \hspace{1cm} \cos \phi' = g'/g'_1 \hspace{1cm} \sin \phi' = g'/g'_2
 \end{align}
and $\tan \beta = v_2/v_1$ is the ratio between the vacuum expectation values of the two Higgs doublets 
of the model.
The analytical expressions show that the corrections to precision data only mildly depends on $\beta$.
We thereby assume $\cos4\beta=0$ and report the resulting best fits in fig.\fig{LH6}a.
Like in the previous model the $M_W$ anomaly can be reproduced,
and collider bounds on vectors and on $W,Y$ can be avoided for large enough gauge couplings.

 \bigskip
 
A related SU(6) little-Higgs model is obtained if only $\SU(2)_1\otimes \SU(2)_2\otimes{\rm U}(1)_Y$ is gauged.
The model is dubbed `incomplete' because the Higgs mass receives quadratically divergent corrections associated to the small $g'$ coupling.
There is no extra $Y'$ vector, no correction to the $Y$ parameter, and
a contribution to $\widehat{T}$ arises because the two different Higgs vacuum expectations break isospin:
\beq\begin{array}{ll}\displaystyle
   \widehat S = \frac{2 M_W^2}{g^2 f^2} \cos^2 \phi,\qquad 
  \label{eq:S2}  \qquad & \displaystyle
  W  = \frac{4 M_W^2}{g^2 f^2} \cos^4 \phi , \\  \displaystyle
   \widehat T  =\frac{M_W^2}{g^2 f^2} \cos^2 2 \beta, &
   \displaystyle
   Y  =0.
  \end{array}\eeq
The resulting best fits are reported in fig.\fig{LH6}b.
Again, the $M_W$ anomaly can be reproduced,
and large $\SU(2)$ couplings are here needed to avoid collider bounds and bounds on $W$.

We verified that
other little-higgs models that do not contribute to $\widehat T$ at tree-level
do not provide good fits to the $M_W$ anomaly.

\subsection{Gauge bosons and Higgs in extra dimensions}
Finally, we recall that similar structures (with the two copies of electroweak vectors replaced by an infinite massive tower)
arise in models with extra dimensions.
A dominant correction to $\widehat T$ is obtained if the Higgs doublet propagates in the extra dimensions
more than the SM vector bosons and the SM fermions.
This can be achieved, for example, considering 
one flat extra dimension with length $1/f$, with the SM fermions confined on 
one boundary, in the presence of vector kinetic terms localized on the boundary~\cite{hep-ph/0405040}:
\begin{equation}
\widehat S=\frac{2}{3} \frac{M^2_W}{f^2}\ ,\qquad
\widehat T=\frac{M_W^2}{3c'f^2} ,\qquad
W=\frac{c M_W^2}{3f^2}\ ,\qquad
 Y=\frac{c' M_W^2}{3f^2}.
\label{eq:extrad}
\end{equation}
The parameters $c$ and $c'$ control the relative size of kinetic terms localized on the boundary for $\SU(2)_L$ and
$\U(1)_Y$ vectors respectively.
A dominant correction to the $\widehat T$ parameter arises for small $c'$, corresponding to mostly-localized 
$\U(1)_Y$ vectors.
Since the theory is non-renormalizable, we cannot claim that eq.\eq{extrad} allows to fit the $M_W$ anomaly compatibly with collider bounds.

\section{Conclusions}\label{concl}
We performed a global fit to electroweak data, finding that the $M_W$ anomaly claimed by the CDF collaboration could be due
to a universal new-physics correction $\widehat T\approx 10^{-3}$
to the $\widehat T$ parameter, corresponding to an effective $|H^\dagger D_\mu H|^2$ operators suppressed by about 
$5\TeV$, if the only $W$-mass measurement included in the global fit is the new CDF result.

Best-fit regions shown in fig.\fig{ST} exhibit a significant correlation of $\widehat T$ with the $\widehat S$ parameter, that can thereby be also present at 
a comparable level.
On the other hand, LHC data now restrict the universal $W,Y$ parameters to values too small to reproduce the $M_W$ anomaly.
This kind of effects could be produced as follows.
\begin{itemize}
\item New physics that gives {\bf tree-level} corrections can have multi-TeV masses, and thereby can easily be compatible
with collider bounds. In section~\ref{treeloop} we discussed scalars with vacuum expectation values.
In section~\ref{Z'} we classified $Z'$ vectors, 
showing that a contribution to $\widehat T$ only can be provided by a $Z'$ coupled to the Higgs only.
In view of the correlation discussed above, various $Z'$ coupled to SM fermions also provide fits with comparable quality, as shown in fig.\fig{Z'}.
Specific little-Higgs models proposed in the literature contain heavy vectors that can fit the $M_W$ anomaly, such as
those based on two copies of SM vectors and
SU(5)/SO(5) and SU(6)/Sp(6) global symmetries and discussed in section~\ref{LH}, with global fits shown in fig.\fig{LH} and\fig{LH6} respectively.
We mention how specific higher dimensional geometries provide similar effects, as a tower of extra vectors.

\item New physics that gives {\bf loop-level} corrections needs to be in a few hundred GeV range, and thereby is easily excluded
by collider bounds. Possible exceptions involve special kinematical configurations, such as a quasi-degenerate
set of particles that decay invisibly into a neutral state, possible DM candidate.
\end{itemize}

\small
\paragraph{Acknowledgments}
We thank Mohamed Aghaie, Alessandro Dondarini, Luca di Luzio, Christian Gross, 
Daniele Teresi, Andrea Tesi, Riccardo Torre, Andrea Wulzer for useful discussions.

\end{document}